\documentclass[12pt]{article}
\usepackage{amsthm,amssymb,amsmath}
\usepackage[american,british]{babel}

\newtheorem*{Th}{Theorem}
\newtheorem*{pro}{Proposition}

\newcommand{\Pf}{\operatorname{Pf}}

\renewcommand{\t}{{\operatorname{t}}}
\newcommand{\I}{\operatorname{i}}

\newcommand{\diag}{\operatorname{diag}}

\newcommand{\C}{{\mathbb C}}
\newcommand{\R}{{\mathbb R}}

\newcommand{\D}{{\partial}}

\begin{document}

\title{Pfaffian Solutions for the\\ Manin-Radul-Mathieu SUSY
KdV\\ and SUSY sine-Gordon Equations}

\author{Q. P. Liu$^1$\thanks{On leave of absence from
Beijing Graduate School, CUMT, Beijing 100083, China}
\thanks{Supported by {\em Beca para estancias temporales
de doctores y tecn\'ologos extranjeros en
Espa\~na: SB95-A01722297}}
   $\,$ and Manuel Ma\~nas$^{2,1}$\thanks{Partially supported by CICYT:
 proyecto PB95--0401}\\
$^1$Departamento de F\'\i sica Te\'orica II,\\ Universidad
Complutense,\\ E28040-Madrid, Spain.\\ $^2$Departamento de
Matem\'atica Aplicada y Estad\'\i stica,\\ Escuela Universitaria de
Ingenier\'\i{}a T\'ecnica Areona\'utica,\\ Universidad
Polit\'ecnica de Madrid,\\ E28040-Madrid, Spain.}
\date{}

\maketitle

\begin{abstract}
We reduce the vectorial binary Darboux transformation for the
Manin-Radul supersymmetric KdV system in such a way that it preserves the
Manin-Radul-Mathieu supersymmetric KdV equation reduction. Expressions  in terms of bosonic
Pfaffians are provided for transformed solutions and wave
functions. We also consider the implications of these results for
the supersymmetric sine-Gordon equation.
\end{abstract}
\newpage

\section{Introduction}

The supersymmetric version of the Korteweg-de Vries (KdV) system
was introduced by Manin and Radul in \cite{mr}. Thereafter many
integrable equations have been extended in this way. The role of
the KdV equation  in two dimensional quantum gravity lead the group
of Alvarez-Gaum\'e to search for analogous structures for
supersymmetric two dimensional quantum gravity \cite{ag1}. Their
results indicated that the supersymmetric extensions of the KdV
equation might be relevant in the study of SUSY 2d quantum gravity.

The Manin-Radul SUSY KdV has a distinguished reduction
\cite{mr,mathieu}, a single equation that we shall call
Manin-Radul-Mathieu SUSY KdV. This equation is closely related to
the SUSY sine-Gordon equation \cite{ik}. In this note we extend to
these equations the Darboux transformations, providing in this
manner efficient ways to construct explicit solutions of these two
equations.  Our scheme is a suitable reduction of the one proposed
in \cite{liu,lm1} (see also \cite{lm2} for super Wro\'nski
determinant soutions), that leads us to express the new solutions
in terms of bosonic Pfaffians.

The layout of the paper is as follows. First, in \S 2, we recall
the reader the basic facts regarding the vectorial binary Darboux
transformation for the Manin-Radul SUSY KdV system, then, in \S 3,
we reduce these results to the Manin-Radul-Mathieu SUSY KdV
equation. Finally, in \S 4, we conclude the letter by applying
these techniques to the SUSY sine-Gordon equation. Let us remark
that for each case we give explicit examples, namely a supersoliton
for the
 Manin-Radul-Mathieu SUSY KdV and a superkink solution
 of the SUSY sine-Gordon equation ---these examples are obtained by
 dressing the zero solution. Obviously, multi-supersoliton and
 multi-superkink solutions are immediately constructed within
 our scheme using the mentioned bosonic Pfaffians.

\section{Vectorial Binary
 Darboux Transformations for the Manin-Radul supersymmetric KdV system}

The MR SUSY KdV system is defined in terms of three independent
variables $\vartheta,x,t$, where $\vartheta\in\C_{\text{a}}$ is an
odd supernumber, and $x,t\in\C_{\text{c}}$ are even supernumbers,
and two dependent variables $\alpha(\vartheta,x,t),
u(\vartheta,x,t)$, where $\alpha$ is an odd function taking values
in $\C_{\text{a}}$ and $u$ is even function with values in
$\C_{\text{c}}$. A basic ingredient is a superderivation defined by
$D:=\partial_{\vartheta}+\vartheta \partial_x$. The system is
\begin{equation}\label{mr}
\begin{aligned}
\alpha_t&={1\over 4} (\alpha_{xxx}+3(\alpha D\alpha)_x+6(\alpha u)_x),\\
u_t&={1\over 4}(u_{xxx}+6uu_x+3\alpha_xDu+3\alpha (Du_x)),
\end{aligned}
\end{equation}
where we use the notation
$f_x:=\partial f/\partial x$ and $f_t:=\partial f/\partial t$.

Let  $E$ be  a supervector space
 over $\boldsymbol \Lambda:=\C_{\text{a}}\oplus \C_{\text{c}}$
    and  $\ell$  an even linear operator over $E$; then,
the linear system
\begin{equation}\label{veclin}
\begin{aligned}
 \psi_{xx}+\alpha D\psi+u\psi-\ell\psi&=0,\\
 \psi_t-{1\over2} \alpha (D\psi_x) -\ell\psi_x-{1\over 2}u\psi_x+
{1\over 4}\alpha_x D\psi
+{1\over 4}u_x\psi&=0,
\end{aligned}
\end{equation}
has as its compatibility condition the MR SUSY KdV system
\eqref{mr}. Eqs. \eqref{mr} are also the compatibility condition of
adjoint linear system:
\begin{equation}\label{aveclin}
\begin{aligned}
\psi^*_{xx}+D(\alpha \psi^* )+u\psi^* -\psi^* m&=0,\\
\psi^*_t+{1\over 2}\alpha D\psi^*_x-\psi^*_x m-{1\over 2}
(u+D\alpha )\psi^*_x +{1\over 4}D(\alpha_x\psi^*)+{1\over
4}u_x\psi^*&=0,
\end{aligned}
\end{equation}
where $\psi^*(\vartheta,x,t)\in\tilde{{E}}^*$ is a linear function
on the supervector space $\tilde{{E}}$, and $m $ is an even linear operator.

In order to construct Darboux transformations for these linear
systems we need to introduce a linear operator, say
$\Omega[\psi,\psi^*]: \tilde E\to E$, bilinear in $\psi$
and $\psi^*$, defined by the compatible equations
\begin{equation}\label{pont}
\begin{aligned}
D\Omega[\psi,\psi^*]=&\psi\otimes \psi^*, \\
\Omega[\psi,\psi^*]_t=&\ell
\Omega[\psi,\psi^*]_x +\Omega[\psi,\psi^*]_xm \\
&- D(\psi_x\otimes\psi^*_x+{1\over
2}uD\Omega[\psi,\psi^*])
-{1\over 4}\alpha_x D\Omega[\psi,\psi^*]\\
&-{1\over2}(D\psi)\otimes((D\alpha)\psi^*-\alpha
(D\psi^*))+{1\over2}\alpha(\psi\otimes \psi^*_x -\psi_x\otimes
\psi^* )
\end{aligned}
\end{equation}
such that
\begin{equation}\label{constraint}
\ell \Omega[\psi,\psi^*]- \Omega[\psi,\psi^*]m =D(\psi_x\otimes\psi^*-\psi\otimes \psi^*_x)-
\alpha \psi\otimes \psi^*.
\end{equation}

Now we are ready to present the following:
\newtheorem*{binary}{\textit{The Vectorial Binary Darboux Transformation
\cite{lm1,lm3}}}

\begin{binary}\label{binary}
Let $\xi(\vartheta,x,t)$ be an even vector
of the supervector space $V$, of total dimension $N$,
satisfying Eq. \eqref{veclin},
$\xi^*(\vartheta,x,t)$  an odd linear functional
of the dual supervector space $V^*$
solving Eq. \eqref{aveclin} and
$\Omega[\xi,\xi^*]$  a non singular even linear operator
on $V$, $\det\Omega[\xi,\xi^*]_{\operatorname{body}}\neq 0$, defined in terms of the
compatible Eqs. \eqref{pont} and \eqref{constraint}. Then, the
objects
\begin{align*}
&\hat{\psi}:=\psi-\Omega[\psi,\xi^*]\Omega[\xi,\xi^*]^{-1}\xi,\\
&\hat{\psi^*}:=\psi^*-\xi^*\Omega[\xi,\xi^*]^{-1}\Omega[\xi,\psi^*],\\
&\hat\alpha=\alpha-2D^3\ln\det \Omega[\xi,\xi^*],\\ &\hat
u=u+2\hat\alpha D\ln\det \Omega[\xi,\xi^*]+2\left(
\frac{\sum_{j=1}^N D(\xi)_j\;\det \Omega[\xi,\xi^*]_j}{\det \Omega[\xi,\xi^*]}\right)_x,
\end{align*}
where  $\Omega[\xi,\xi^*]_j$ is an operator with associated
supermatrix obtained from the corresponding one of
$\Omega[\xi,\xi^*]$ by replacing the $j$-th column by $\xi$,
satisfy the Eqs. \eqref{veclin} and \eqref{aveclin} whenever
the unhatted variables do. Thus, $\hat\alpha$ and $\hat u$ are new
solutions of \eqref{mr}.
\end{binary}

Observe that here we are using ordinary determinants, this is possible because
we are considering only even matrices, so that its coefficients conmute,
and also because we have finite total dimension $N$.

\section{The reduction to the Manin-Radul-Mathieu supersymmetric KdV equation}

The equations \eqref{mr} admits the  reduction, $u=0$, due to Manin
and Radul \cite{mr} and  studied later by Mathieu \cite{mathieu}.
The system reduces then to the single equation
\begin{equation}\label{mathieu}
4\alpha_t=\alpha_{xxx}+3(\alpha D\alpha)_x,
\end{equation}
the so called Manin-Radul-Mathieu supersymmetric KdV  equation.

 Obviously this equation is the compatibility condition of
the following linear system
\begin{equation}\label{linmathieu}
\begin{aligned}
 \psi_{xx}+\alpha D\psi-\ell\psi&=0,\\
 \psi_t-{1\over2} \alpha (D\psi_x) -\ell\psi_x+
 {1\over 4}\alpha_x D\psi&=0.
\end{aligned}
\end{equation}

We first observe that given an even solution $\xi$ of
\eqref{veclin} then $D\xi^\t$ is an odd solution of \eqref{aveclin}
if and only if $Du=0$. Second, if $u=u_0$, a constant, given $\xi$
and $\psi$ solutions of \eqref{veclin}  we can take $\xi^*=D\xi^\t$
and $\psi^*=D\psi^\t$ as solutions of \eqref{aveclin}   whenever
$m=\ell^\t$, then
\[
\Omega[\psi,\xi^*]+\Omega[\xi,\psi^*]^\t-\psi\otimes\xi^t
\]
is a constant linear operator.

Now, if we perform the vectorial binary Darboux transformation for
the MR SUSY KdV system induced by the above transformation data and
impose
\begin{equation}\label{constraints}
\begin{aligned}
\Omega[\xi,\xi^*]+\Omega[\xi,\xi^*]^\t&=\xi\otimes\xi^t,\\
\Omega[\psi,\xi^*]+\Omega[\xi,\psi^*]^\t&=\psi\otimes\xi^t,
\end{aligned}
\end{equation}
then, the transformed wave functions $\hat\psi$ and $\hat\psi^*$
are linked by $\hat\psi^*=D\hat\psi^\t$, so that $D\hat u=0$.
Moreover, one can show that $\hat u=u_0$, so if $u=0$ then $\hat
u=0$.

We can summarize these results in the following
\begin{pro}
Given solutions $\psi\in \C_{\operatorname{c}}$ and $\xi$,
an even vector in $V$, of
\eqref{linmathieu} and potentials subject to
\eqref{constraints}; then, the vectorial binary Darboux
transformation preserves the Manin-Radul-Mathieu supersymmetric KdV equation reduction.
\end{pro}

At this point one could apply the Grammian type expressions of the
vectorial binary Darboux transformation of the MR SUSY KdV system
to get new solutions of the MRM SUSY KdV equation from given ones.
However, these expressions for the components of the wave functions
and fields can be rewritten compactly in terms of ordinary
Pfaffians, the derivation of this results follows the lines given
in \cite{lm4}. But first let us remind the reader what is a
Pfaffian of an even-dimensional skew-symmetric matrix.  If
$S=(s_{ij})_{i,j=1,\dots,2M}$ is an even-dimensional skew matrix,
$s_{ij}+s_{ji}=0$, its Pfaffian  is defined as
\[
\Pf S=\sum_{\pi\in P}\operatorname{sgn}(\pi)\prod_{k=0}^{M-1}
s_{\pi(2k+1)\pi(2k+2)}
\]
where $P$ is the set of permutations $\pi$ of the $2M$ first natural numbers
 such that
\begin{enumerate}
\item For $i,j=0,\dots,M-1$, with $i<j$, then $\pi(2i+1)<\pi(2j+1)$.
\item For any $j=1,\dots,M$ one has $\pi(2j-1)<\pi(2j)$.
\end{enumerate}
For example, for $M=2$ we have $\Pf
S=s_{12}s_{34}-s_{13}s_{24}+s_{14}s_{23}$. An important property of
Pfaffians is that for any matrix $M$ we have $\Pf(M^\t
SM)=\det(M)\Pf S$, from where it follows that $\det S=(\Pf S)^2$.
Obviously, as we already did with ordinary determinants, we can
extend this construction to have Pfaffians of
 even linear operators on supervector spaces with even total dimension, $N=2M$.

Notice that the potential matrices $\Omega[\xi,\xi^*]$ and
$\Omega[\xi,\xi^*]$ split  as follows
\begin{align}
\label{S1}\Omega[\psi,\xi^*]&=\frac{1}{2}\big[\xi\otimes\xi^\t+
S\big],\\
\label{S2}\Omega[\psi,\xi^*]&=\frac{1}{2}\big[\psi\xi^\t+
S[\psi,\xi]^\t\big],
\end{align}
where $S$ is skew-symmetric and satisfy
\begin{align*}
D S&=\xi\otimes D\xi^\t-(D \xi)\otimes \xi^\t,\\ D
S[\psi,\xi]&=\psi\D\xi-(D \psi) \xi.
\end{align*}
In terms of $S$ the constraint \eqref{constraint} reads
\begin{equation}\label{constraintS}
\ell S-S\ell^\t=
2\big[(D\xi_x)\otimes(D\xi^\t)-(D\xi)\otimes(D\xi_x^\t)+
\xi_x\otimes\xi_x^\t\big]-\ell\xi\otimes\xi^\t-\xi\otimes\xi^\t\ell^\t.
\end{equation}

\newtheorem*{vectop}{\textit{
Pfaffian Form of the Reduced
 Transformation}}

\begin{vectop}
The reduction of the Grammian determinant type solutions of the MR
SUSY KdV system to the MRM SUSY KdV equation gives
\begin{enumerate}
\item For $N$ even
\begin{align*}
  \hat \alpha &= \alpha-4 D^3(\ln\Pf(S)),\\
\hat\psi&=\frac{
\Pf \begin{pmatrix}     0&\psi & S[\psi, \xi]^\t\\
                   -\psi & 0   &-\xi^\t\\
                   -S[\psi,\xi]&\xi&S
                   \end{pmatrix}}{\Pf(S)}.
\end{align*}
\item For $N$ odd
\begin{align*}
  \hat\alpha &=\alpha-4 D^3\Big(\ln\Pf\begin{pmatrix} 0 &-\xi^\t\\
\xi & S\end{pmatrix}\Big),\\[.2cm]
\hat\psi&=\frac{\Pf\begin{pmatrix} 0&S[\psi,\xi]^\t\\
                 -S[\psi,\xi]&S\end{pmatrix}}
                 {\Pf\begin{pmatrix}0& -\xi^\t\\
                 \xi&S\end{pmatrix}}.
\end{align*}
\end{enumerate}

\end{vectop}

If $\ell=\diag(\ell_1,\dots,\ell_N)$ is a diagonal matrix with
$\ell_i\in\C_{\operatorname{c}}$ all different, $\ell_i\neq \ell_j$
for $i\neq j$, then \eqref{constraintS} imply that the components
of $\xi$, $\xi_i$, $i=1,\dots, N$ are subject to
\begin{equation}\label{constraintxi}
 2(D\xi_i)_x(D\xi_i)
+{(\xi_i)_x}^2-\ell_i{\xi_i}^2=0
\end{equation}
and determines $S=(s_{ij})$  as
\[
s_{ij}=\frac{2}{\ell_i-\ell_j}\big[
(D\xi_i)_x(D\xi_j)-(D\xi_i)(D\xi_j)_x
+(\xi_i)_x(\xi_j)_x-(\ell_i+\ell_j)\xi_i\xi_j
\big].
\]
In particular, for zero background $\alpha=0$,
\[
\xi_i(\theta,x,t)=(a_{i,0}^++\theta a_{i,1}^+)\exp(k_ix+k_i^3 t)+
(a_{i,0}^-+\theta a_{i,1}^-)\exp(-k_ix-k_i^3 t),
\]
where $k_i^2=\ell_i$, $a_{i,0}^\pm$ and  $a_{i,1}^\pm$ even and odd
supernumbers, respectively, and
\eqref{constraintxi} simply is
\[
a_{i,1}^+a_{i,1}^-=k_ia_{i,0}^+a_{i,0}^-.
\]
For example, the solution for $N=1$ is \cite{lm3}
\[
\hat\alpha=\alpha-4\frac{(D\xi_x)\xi-\xi_x(D\xi)}{\xi^2},
\]
that for $\alpha=0$ gives
\[
\hat\alpha=-\frac{8k(a_1^+a_0^--a_1^-a_0^+)}{
(a_0^+\exp(\eta)+a_0^-\exp(-\eta))^2}\Big[
1-2\theta\frac{a_1^+\exp(\eta)+a_1^-\exp(-\eta)
}{a_0^+\exp(\eta)+a_0^-\exp(-\eta)}
\Big],
\]
where $\eta(x,t):=kx+k^3t$ and we are assuming that the even
quantity $a_0^+\exp(\eta)+a_0^-\exp(-\eta)$ is invertible. Which
can be considered as the typical 1-soliton of the MRM SUSY KdV
equation; hence, our Pfaffian solutions, when applied to the zero
background, provides the multi-soliton solutions of this equation.

\section{Pfaffian solutions of the supersymmetric sine-Gordon equation}

The SUSY sine-Gordon equation \cite{ik} is
\begin{equation}\label{sg}
DD_t\Phi=\sin(\Phi)
\end{equation}
where $D_t=\dfrac{\partial}{\partial \theta_t}+
\theta_t\dfrac{\partial}{\partial t}$ and now $\theta,\theta_t\in\R_{\operatorname{a}}$ and
$\Phi,x,t\in\R_{\operatorname{c}}$.
 The linear
system
\
\begin{equation}
 \label{lsg}
\begin{aligned}
  \psi_x & =\I(D\Phi)D\psi+K \bar\psi \\
 D_t\psi & =-\dfrac{1}{2}K^{-1}\exp(\I\Phi)D\bar\psi,
\end{aligned}
\end{equation}
has as their compatibility \eqref{sg}, here $\psi$  is an even
vector in a supervector space and $K$ an invertible real even
operator over it.
 It can be easily shown  that
\[
\psi_{xx}+\alpha D\psi-\ell\psi=0
\]
where  $\alpha:=\gamma D\gamma-\I\gamma_x$
 with $\gamma=D\Phi$ an odd real supervariable and $\ell=K^2$.

A standard application \cite{wadati,nimmo} of the results of the
previous section yields

\begin{Th}
Let $\xi$ be an even vector
solution in the  supervector space $V$, of total dimension $N$,
 of \eqref{lsg} so that
\eqref{constraintS} are satisfied; then,
\[
\hat\Phi-\Phi=\begin{cases}
4\operatorname{arg}\Pf S,&\text{ when $N$ is even,}\\[.4cm]
4\operatorname{arg}\Pf\begin{pmatrix} 0 &-\xi^\t\\
\xi &  S\end{pmatrix},
&\text{ when $N$ is odd.}
\end{cases}
\]
is a new solution of (\ref{sg}).
\end{Th}
When $K=\diag(k_1,\dots,k_N)$ and $\Phi=0$ the solution of \eqref{lsg} has components
\begin{multline*}
\xi_i=\Big[
\big(1-\frac{1}{2}\theta_t\theta\big)A_{i,0}^++\big(\theta-\frac{1}{2k_i}\theta_t\big)
A_{i,1}^+\Big]\exp(\eta_i(x,t))\\+\I
\Big[
\big(1-\frac{1}{2}\theta_t\theta\big)A_{i,0}^-+\big(\theta+\frac{1}{2k_i}\theta_t\big)
A_{i,1}^-\Big]\exp(-\eta_i(x,t))
\end{multline*}
where now
\[
\eta_i(x,t):=k_ix-\frac{1}{4k_i}t,\quad \quad
 A_{i,1}^+A_{i,1}^-=k_iA_{i,0}^+A_{i,0}^-.
\]
For $N=1$ we get the superkink solution
\[
\hat\Phi=4\arctan\Bigg(
\frac{\big(1-\dfrac{1}{2}\theta_t\theta\big)A_0^-+\big(\theta
+\dfrac{1}{2k_i}\theta_t\big)
A_1^-}{\big(1-\dfrac{1}{2}\theta_t\theta\big)A_0^+
+\big(\theta-\dfrac{1}{2k_i}\theta_t\big)
A_1^+}\exp\Big(-2kx+\frac{1}{2k}t\Big)
\Bigg),
\]
if we assume that $A_0^+\in\R_{\operatorname{c}}$ is  invertible
 we can write
\begin{multline*}
\Phi=4\arctan\bigg(\frac{1}{A_0^+}\Big(
A_0^-+\theta\dfrac{A_1^-A_1^+-A_1^+A_0^+}{A_0^+}-
\theta_t\dfrac{A_1^-A_1^++A_1^+A_0^+}{2 k A_0^+}
\\-\theta_t\theta A_0^-\Big)
\exp\Big(-2kx+\frac{1}{2k}t\Big)
\bigg).
\end{multline*}

The formulae of our theorem obviously contains the supersymmetric
extension of topological solitons like the multi-kinks solutions.

\end{document}